\documentclass[superscriptaddress,showpacs,preprintnumbers,amsmath,amssymb,prl,twocolumn]{revtex4}

\usepackage{graphicx}
\usepackage{dcolumn}
\usepackage{bm}
\newcommand{\kb}{k_{\mbox{\tiny B}}}
\renewcommand{\vec}[1]{\bm{#1}}

\begin{document}

\title{\textit{Ab initio} Green-Kubo Approach\\ for the Thermal Conductivity of Solids}

\author{Christian Carbogno}%
\affiliation{%
Fritz-Haber-Institut der Max-Planck-Gesellschaft, Faradayweg 4-6, D-14195 Berlin, Germany
}%
\author{Rampi Ramprasad}%
\affiliation{%
University of Connecticut, 97 North Eagleville Road, Storrs, Connecticut 06269, USA
}
\author{Matthias Scheffler}%
\affiliation{%
Fritz-Haber-Institut der Max-Planck-Gesellschaft, Faradayweg 4-6, D-14195 Berlin, Germany
}%
\affiliation{Department of Chemistry and Biochemistry, University of California at Santa Barbara, CA 93106, USA}
\affiliation{Materials Department, University of California at Santa Barbara, CA 93106, USA}
\date{\today}

\begin{abstract}
We herein present a {\it first-principles} formulation of the {\it Green-Kubo} method that allows the accurate
assessment of the phonon thermal conductivity of solid semiconductors and insulators in equilibrium {\it ab initio molecular dynamics} 
calculations. Using the virial for the nuclei, we propose a unique {\it ab initio} definition of the heat flux.
Accurate size- and time convergence are achieved within moderate computational effort by a robust, asymptotically
exact extrapolation scheme. We demonstrate the capabilities of the technique by investigating the thermal conductivity of 
extreme high and low heat conducting materials, namely diamond Si and tetragonal ZrO$_2$.
\end{abstract}


\maketitle

Macroscopic heat transport is an ubiquitous phenomenon in condensed matter that plays a crucial role in a multitude of 
applications,~e.g.,~energy conversion, catalysis, and turbine technology. 
Whenever a temperature gradient~$\nabla T(\vec{R})$ is present, a heat flux~$\vec{J}(\vec{R})$ 
spontaneously develops to move the system back toward
thermodynamic equilibrium.
The temperature- and pressure-dependent thermal conductivity~$\vec{\kappa}(T,p)$ 
of the material
describes the proportionality between heat flux and temperature gradient~(Fourier's law)
\begin{equation}
\vec{J}(\vec{R}) = - \vec{\kappa}(T,p)\cdot\nabla T(\vec{R}) \;.
\label{Fourier}
\end{equation}
In insulators and semiconductors, the dominant contribution to~$\vec{\kappa}(T,p)$ stems from the
vibrational motion of the atoms~(phonons)~\cite{AshcroftMermin}. In spite of significant 
efforts, 
a parameter-free, accurate theoretical approach that allows to assess the thermal conductivity tensor both in the case of weak and strong anharmonicity is still lacking:
Studies of model systems via  classical molecular dynamics~(MD) based on force fields~(FF) can unveil general rules and 
concepts~\cite{Donadio:2009fo}.
However, the needed accuracy for describing anharmonic effects is often not correctly captured by FFs~\cite{Broido:2005kf} and 
trustworthy FFs are generally not available for ``real'' materials used in scientific and industrial applications. 

Naturally, first-principles electronic-structure theory lends itself to overcome this deficiency by
allowing a reliable computation of the inter-atomic interactions. However, severe 
limitations affect the approaches that have hitherto been employed in {\it ab initio} frameworks 
for studying the thermal conductivity of solids: 
(a)~Approaches based on the~{\it 
Boltzmann Transport Equation}~\cite{Broido:2007ua,deKoker:2009cr,Tang:2010vu,Garg:2011hi} account 
for the leading, lowest order contributions to the anharmonicity. Accordingly, these approaches are
justified at low temperatures, at which they also correctly describe relevant nuclear quantum effects.
At elevated temperatures and/or in the case of strong anharmonicity, this approximation is however known to break down~\cite{Ladd:1986tv,Turney:2009ur}.
(b)~{\it Non-Equilibrium} approaches~\cite{Gibbons:2009hd,Gibbons:2011je,Stackhouse:2010vg}
require to impose an artificial temperature gradient, which becomes unreasonably large~($\gg 10^9$~K/m)
in the limited system sizes accessible in {\it first-principles} calculations. Especially at high 
temperatures, this can lead to non-linear artifacts~\cite{Tenenbaum:1982wu,Schelling:2002jl,He:2012hg} that
prevent the assessment of the linear response regime described by Fourier's law.

In this letter, we present an {\it ab initio} implementation of the {\it Green-Kubo}~(GK) method~\cite{Kubo:1957tf},
which does not suffer from the aforementioned limitations~\cite{Schelling:2002jl,Broido:2007ua}, 
since $\vec{\kappa}(T,p)$ is determined from {\it ab initio} molecular dynamics simulations~(aiMD) 
in thermodynamic equilibrium that account for anharmonicity to all orders. Hitherto, fundamental challenges 
have prevented an application of this technique in a first-principles framework: Conceptually,
a definition of the heat flux associated with vibrations in the solid is required; numerically,
the necessary time- and length scales need to be reached.
First, we succinctly describe how we overcome the conceptual hurdles,~i.e., the unique {\it ab initio} definition
of the microscopic heat flux (and its fluctuations) for solids. Second, we discuss how this allows to overcome
the numerical hurdles by introducing a robust extrapolation scheme, so that time- and size convergence is achieved  within moderate 
computational effort. Third, we validate our formalism and demonstrate its wide applicability by investigating the
thermal conductivity of diamond Si and tetragonal ZrO$_2$~($\mathit{P4_2/nmc}$), two materials that 
feature especially large/low thermal conductivities due to being particularly harmonic/anharmonic.

For a given pressure $p$, volume~$V$, and temperature~$T$, the fluctuation-dissipation theorem,
which is the only assumption entering the GK formula, relates 
the cartesian components~$\alpha\beta$ of the thermal conductivity tensor
\begin{equation}
{\kappa}_{\alpha\beta}(T,p) = \frac{V}{\kb T^2}\lim_{\tau\rightarrow\infty}  \int\limits_0^{\tau} \left\langle {G}[\vec{J}]_{\alpha\beta}(\tau') \right\rangle_{(T,p)} \;  d\tau' 
\label{GreenKubo}
\end{equation}
to the time-(auto)correlation functions 
\begin{equation}
{G}[\vec{J}]_{\alpha\beta}(\tau) =  \lim_{t_0\rightarrow\infty} \frac{1}{t_0}\int\limits_0^{t_0-\tau} {J}_\alpha(t)\; {J}_\beta(t+\tau)\; dt 
\label{AC}
\end{equation}
of the heat flux~$\vec{J}(t)$.
In Eq.~(\ref{GreenKubo}), $\kb$ is the Boltzmann constant and $\left\langle\cdot\right\rangle_{(T,p)}$ denotes an ensemble average that
is performed by averaging over multiple correlation functions~${G}[\vec{J}]_{\alpha\beta}(\tau)$,
which are individually computed from different MD trajectories~(time span~$t_0$, microcanonical ensemble~\cite{MACDOWELL:1999fe}) 
using Eq.~(\ref{AC}).

First, the GK method requires a consistent definition of the heat flux~$\vec{J}(t)$. 
Common FF-based formalisms~\cite{MACDOWELL:1999fe,Hardy:1963gh,Helfand:1960uy} achieve such a definition by 
partitioning the total,~i.e.,~kinetic and potential, energy of the system~$E      = \sum_{I} E_{I}$
into contributions~$E_I$ associated with the individual atoms~$I$. Using their positions~$\vec{R}_I$,
the energy density associated with the nuclei is $e(\vec{R}) = \sum_I E_I \delta(\vec{R}-\vec{R}_I)$ with  the
Delta distribution~$\delta(\vec{R})$.
With this subdivision, the integration of the 
continuity equation~$\partial e(\vec{R},t)/\partial t + \nabla \cdot \vec{j}(\vec{R},t) = 0$ for the heat flux density~$\vec{j}(\vec{R},t)$
reveals that the total heat flux 
\begin{equation}
\vec{J}(t)=\frac{1}{V} \frac{d}{dt}\sum_{I} \vec{R}_I E_I
\label{Bary}
\end{equation}
is related to the motion of the energy barycenter.
Conceptually, the required partitioning is straightforward for FFs and 
challenging in a first-principles framework~\cite{Yu:2011kv,Marcolongo:2016dn}, but in
neither of the cases unique~\cite{Howell:2012fh,Marcolongo:2016dn}.
Using a combined nuclear and electronic energy density, Marcolongo, Umari, and Baroni 
recently proposed a non-unique formulation of the heat flux and used it to study the convective heat flux in liquids from first principles~\cite{Marcolongo:2016dn}. 
However, their approach is numerically unsuitable for the conductive thermal transport in solids, which features much longer lifetimes and mean 
free paths. 
We overcome this limitation by disentangling the different contributions to the heat flux 
and thus finding a unique definition for the conductive heat flux in solids. In
turn, this allows to establish a link to the quasi-particle (phonon) picture of heat transport and thus
to overcome finite time and size effects, as described in the second part of this letter.

For this purpose, we perform the time derivative in Eq.~(\ref{Bary}) analytically~\cite{MACDOWELL:1999fe,Hardy:1963gh}
\begin{eqnarray}
\vec{J}(t) & = & \underbrace{\frac{1}{V}\sum_{I} \dot{\vec{R}}_I E_I}_{\vec{J}_c(t)} + \underbrace{\frac{1}{V}\sum_{I} \vec{R}_I \dot{E_I}}_{\vec{J}_v(t)} \;.
\label{disentangle}
\end{eqnarray}
The first term~$\vec{J}_c(t)$,
which describes convective contributions to the heat flux, requires an energy partitioning scheme, but gives
no contributions to the conductivity tensor in solids~\cite{Howell:2012fh}, as mass transport is negligible. 
Conversely, the dominant virial or conductive term~$\vec{J}_v(t)$ 
does not require an {\it ad hoc} partitioning of the energy~$E$. Only derivatives of the energy~$E_I$
enter~$\vec{J}_v(t)$, so that the forces~$\vec{F}_I=-{\partial U}/{\partial \vec{R}_I}$,~i.e.,~the gradients of
the potential energy surface~$U$, naturally disentangle the individual atomic contributions in a unique fashion  
-- both in FF and \textit{ab initio} frameworks. By rewriting the individual contributions in terms of relative 
distances~$\vec{R}_{IJ} = \vec{R}_I - \vec{R}_J$ we get a definition of the virial flux that is compatible
with periodic boundary conditions
\begin{eqnarray}
\vec{J}_v(t) & = &  -\frac{1}{V}   \sum_{I,J} (\vec{R}_I-\vec{R}_J) (\nabla_{\vec{R}_{IJ}}  U)      \cdot \dot{\vec{{R}}}_I \nonumber\\
             & = &   \sum_{I} \vec{\sigma}_I \cdot \dot{\vec{R}}_I \;.
\label{HFV}
\end{eqnarray}
As discussed in the detailed derivation provided in the Supp. Mat., 
the latter notation highlights that the terms in $\vec{J}_v(t)$ are the individual
atomic contributions~$\vec{\sigma}_I$ to the stress strensor~$\vec{\sigma} = \sum_{I} \vec{\sigma_I}$.

In {\it density-functional theory}~(DFT), the potential-energy surface 
\begin{equation}
U = E^{\text{DFT}} + \frac{1}{2}\sum_{I,J\neq I} \frac{Z_IZ_J}{|\vec{R}_I-\vec{R}_J|} 
\end{equation}
is given by the total energy~$E^{\text{DFT}}$ of the electrons with their ground-state density~$n(\vec{r})$ plus the electrostatic repulsion
between the nuclei with charges~$Z_I$.
Use of the {Hellman-Feynman} theorem leads to a definition for the cartesian components~$\alpha\beta$ of the virials for the individual nuclei 
\begin{eqnarray}
{\sigma}_I^{\alpha\beta} & = &  - \frac{Z_I}{V}\left(\int d\vec{r}\;n(\vec{r}) \frac{({r}^\alpha-{R}_I^\alpha)({r}^\beta-{R}_I^\beta)}{|\vec{r}-\vec{R}_I|^3} \right. \nonumber\\
 && \left. - \frac{1}{2}\sum_{J\neq I} Z_J \frac{({R}_J^\alpha-{R}_I^\alpha)({R}_J^\beta-{R}_I^\beta)}{|\vec{R}_J-\vec{R}_I|^3} \right)\label{sigma}\;,
\end{eqnarray}
whereby all electronic contributions stem from the interaction with the ground state electron density~$n(\vec{r})$.
In turn, this enables a straightforward and unique evaluation of~$\vec{J}_v(t)$ using Eq.~(\ref{HFV}), 
since neglecting the convective term~$\vec{J}_c(t)$ from the beginning allows to integrate out the 
internal electronic contributions to the heat flux~(see Supp. Mat.).  
This holds true also in our practical
implementation of the virial and the analytical stress tensor~\cite{Knuth:2015kc}, since Pulay terms and alike
that can arise can again be associated to individual atoms.
Since both Eq.~(\ref{HFV}) and Eq.~(\ref{sigma}) are exact and non-perturbative, evaluating $\vec{J}_v(t)$ along 
the {\it ab initio} trajectory accounts for the full anharmonicity. 

\begin{figure}
  \centerline{\includegraphics[width=0.85\linewidth]{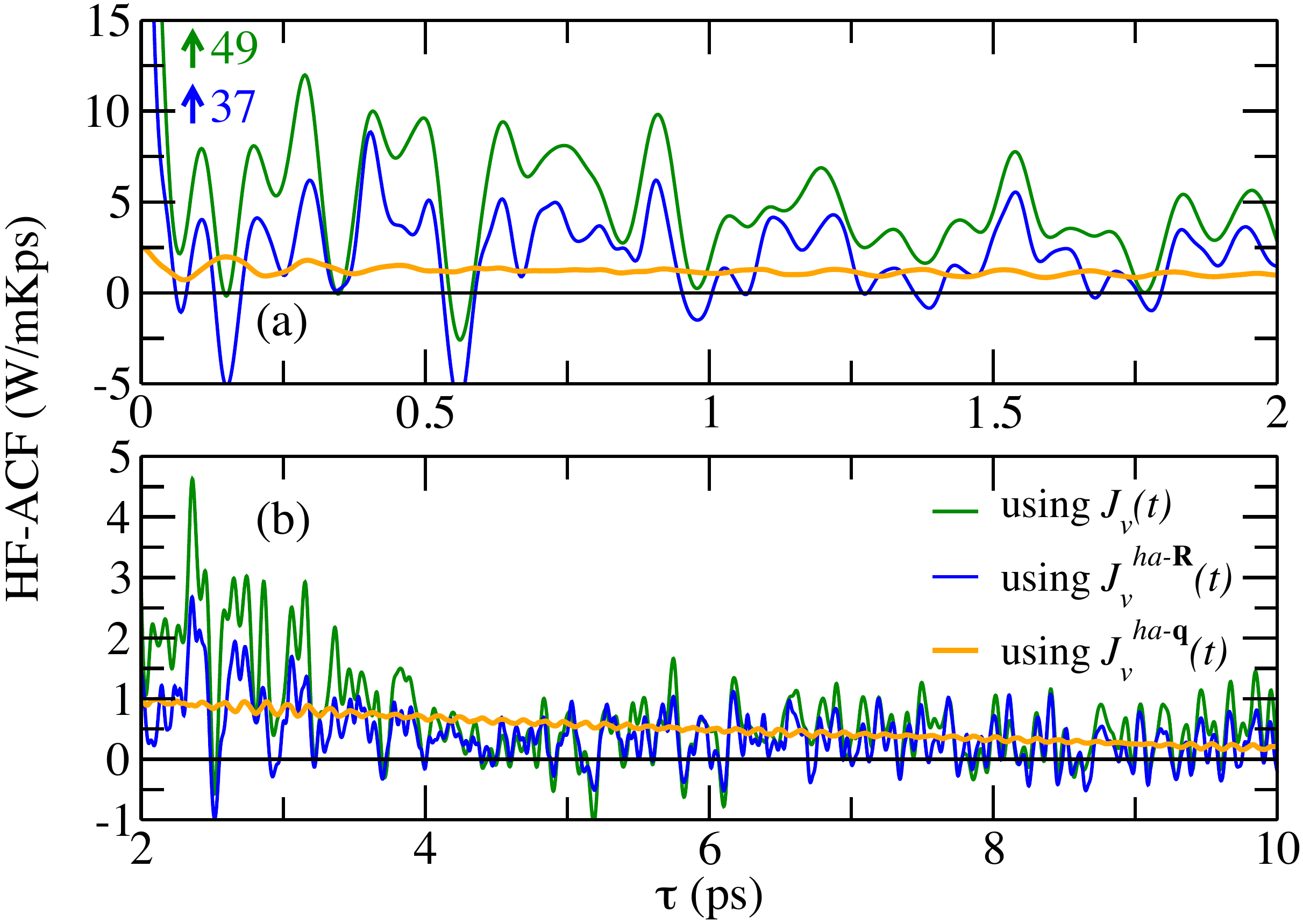}}
  \caption{Early~(a) and late~(b) decay of the heat flux autocorrelation function~(HF-ACF) of Silicon computed in a 64~atom supercell with DFT-LDA at a temperature of 960~K~(trajectory length $\sim207$~ps). 
The green line~(${G}[\vec{J}_v]$) employs the virial {\it ab initio} heat flux~$\vec{J}_v(t)$ that incorporates all anharmonic effects, whereas the blue  and orange 
lines show the HF-ACFs~${G}[\vec{J}_v^{ha}]$ and ${G}[\vec{J}_v^{qp}]$ for approximate heat fluxes computed for the exact same trajectory, but imposing the harmonic approximation,~i.e.,~using~$\vec{J}_v^{ha}(t)$ and~$\vec{J}_v^{qp}(t)$ defined via Eq.~(\ref{VirialHarmonic}) and~(\ref{Peierls}), respectively.}
  \label{ACFSIraw}
\end{figure}

To validate our implementation of the proposed approach in the all-electron, numeric atomic orbital electronic structure code {\it FHI-aims}~\cite{Blum:2009fe} 
we compare the heat flux autocorrelation function~(HF-ACF) computed from first principles~${G}[\vec{J}_v]$
with the respective harmonic HF-ACF~${G}[\vec{J}_v^{ha}]$ by evaluating the approximate virial heat flux~$\vec{J}_v^{ha}(t)$
using the harmonic force constants~$\Phi_{IJ}^{\alpha\beta}=\partial^2 U / \partial{R}_I^\alpha \partial{R}_J^\beta$. 
In the {\it harmonic approximation}, the virials
\begin{equation}
\left({\sigma_I}^{\alpha\beta}\right)^{ha-\vec{R}} = \frac{1}{2V} \sum_{J\neq I}  \Phi_{IJ}^{\alpha\beta}(\Delta {R}_I^\alpha - \Delta {R}_J^\alpha) ({R}_{I}^\beta - {R}_J^\beta) 
\label{VirialHarmonic}
\end{equation}
depend only on the positions and displacements from equilibrium~$\Delta \vec{R}_I=\vec{R}_I-\vec{R}_I^{eq}$~\cite{Ladd:1986tv},
so that $\vec{J}_v^{ha}(t)$ can be evaluated using Eq.~(\ref{HFV}) and~(\ref{VirialHarmonic})
along the exact same first-principles trajectory used to compute~$\vec{J}_v(t)$.
As an example, Fig.~\ref{ACFSIraw} shows such a comparison: ${G}[\vec{J}_v]$
and ${G}[\vec{J}_v^{ha}]$ closely resemble each other and become equal for large time-lags~$\tau$,
which demonstrates the validity of the introduced first-principles definition of the heat flux
and its applicability in {\it ab initio} GK calculations.

However, Fig.~\ref{ACFSIraw} also neatly exemplifies the severe computational challenges of such first-principle
GK simulations: Due to the limited time scales accessible in aiMD runs,
thermodynamic fluctuations dominate the HF-ACF, which in turn prevents a reliable and numerically stable assessment of the 
thermal conductivity via~Eq.~(\ref{GreenKubo}). Furthermore, achieving convergence with respect to system size is 
numerically even more challenging, as classical MD studies based on FFs~\cite{He:2012hg,Howell:2012fh} 
have shown, so that {\it ab initio} GK simulations of solids appear to be computationally prohibitively costly.
However, as we will show below, the computational effort can be reduced by several orders of magnitude
by a correct extrapolation technique employing a proper interpolation in reciprocal space.

\begin{figure}
  \centerline{\includegraphics[width=0.85\linewidth]{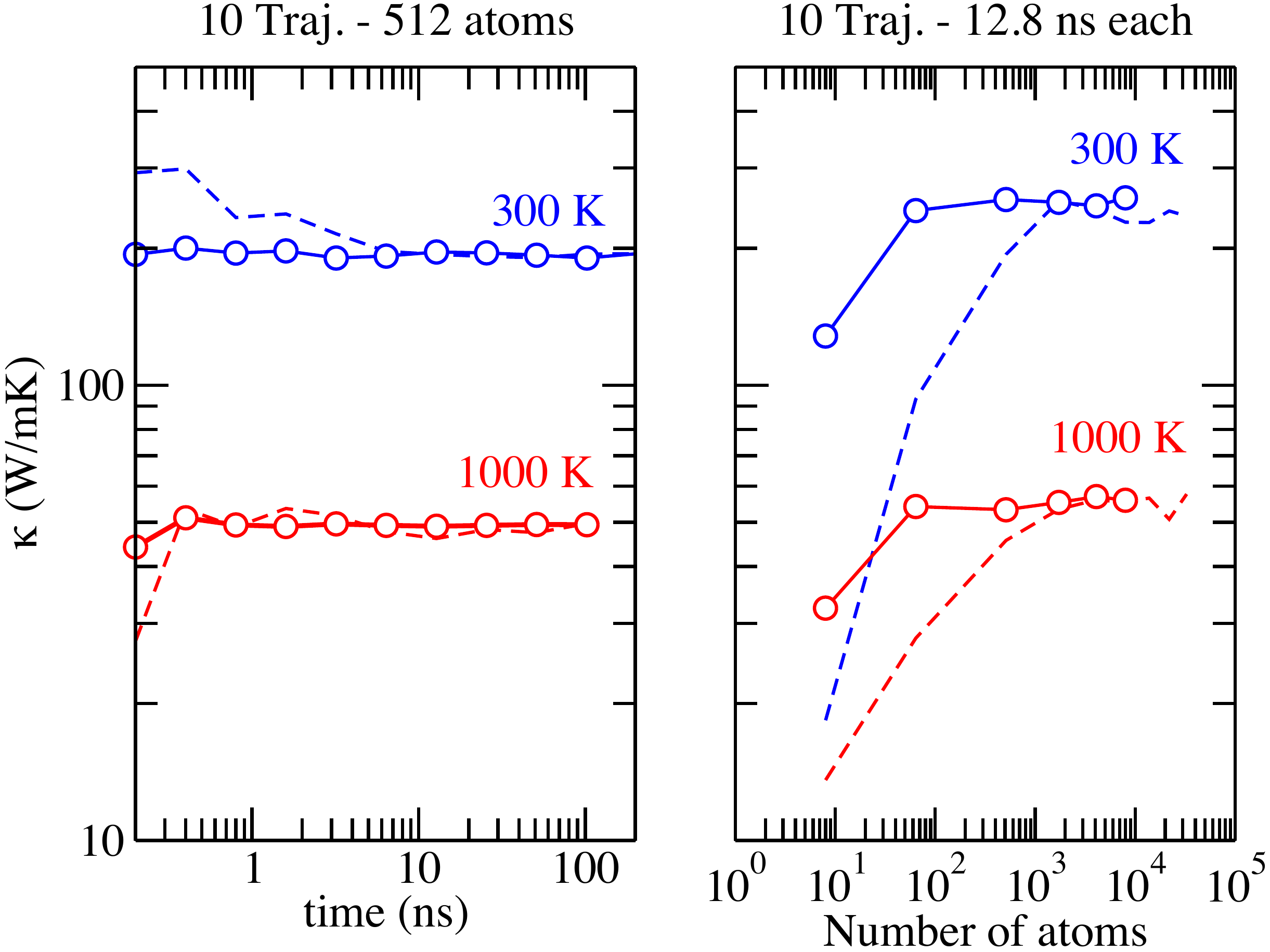}}
  \caption{Thermal conductivity ${\kappa}_{\alpha\alpha}$ of Si at 300 and 1000~K computed with the Tersoff-FF. 
   Values extrapolated (left: in time; right graph: in time and size) with our aiGK method 
   are denoted by circles, whereas the dashed lines show the values resulting 
   from a brute force evaluation of~${G}[\vec{J}_v]$ using only~$\vec{J}_v$,~i.e.,~without any extrapolation. 6912~$\vec{q}$-points corresponding to 
   a $12\times12\times12$~cubic supercell were used for the size extrapolation in the right plot.}
  \label{ConvergenceSi}
\end{figure}

For this purpose, we first note that in the harmonic approximation the HF-ACF can be equivalently~\cite{Ladd:1986tv} evaluated
in reciprocal space using the heat flux definition in the phonon picture~\cite{Peierls:1929jv}:
\begin{equation}
\vec{J}_v^{ha-\vec{q}}(t) = \frac{1}{2V}\sum_{s\vec{q}} n_{s}(\vec{q},t) \; \omega_{s}^2(\vec{q})\; \vec{v}_{s}(\vec{q}) \;.
\label{Peierls}
\end{equation}
Here, the sum goes over all reciprocal space points~$\vec{q}$ commensurate 
with the chosen supercell; $\omega_{s}(\vec{q})$ are the eigenfrequencies and $\vec{v}_{s}(\vec{q})$ are the group 
velocities of the phonon mode~$s$, which are obtained by Fourier transforming and diagonalizing the 
mass-scaled force constant matrix~$\Phi_{IJ}^{\alpha\beta}$ introduced in Eq.~(\ref{VirialHarmonic}).
The time-dependent phonon amplitudes and thus the occupation numbers~$n_{s}(\vec{q},t)$ can be 
extracted from the MD trajectory using the techniques described in Ref.~\cite{McGaughey:2004iz}. 
Accordingly, we can reformulate the HF-ACF as $\hat{{G}} = {G}[\vec{J}_v] - {G}[\vec{J}_v^{ha-\vec{R}}] + {G}[\vec{J}_v^{ha-\vec{q}}]$. 
For fully time and size converged calculations, $\hat{{G}}$ equals ${G}[\vec{J}_v]$;
for underconverged calculations~(cf.~Fig.~\ref{ACFSIraw}), $\hat{{G}}$ exhibits 
significantly less thermodynamic fluctuations, since the  phases of the individual modes do 
not enter Eq.~(\ref{Peierls}). 
 
Even more importantly, this formalism enables a straightforward size-extrapolation 
by extending the sum over (the finite number of commensurate) reciprocal space points~$\vec{q}$ 
in Eq.~(\ref{Peierls}) to a denser grid. The required frequencies~$\omega_{s}(\vec{q}')$ and group velocities~$\vec{v}_{s}(\vec{q}')$ can be
determined on arbitrary $\vec{q}'$-points that are not commensurate with the supercell by Fourier 
interpolating the force constants~$\Phi_{IJ}^{\alpha\beta}$~\cite{Parlinski:1997kr}. In the same spirit, 
we introduce the dimensionless quantity 
\begin{equation}
\Delta{n}_{s}(\vec{q},\tilde{t}) = \frac{n_{s}(\vec{q},t = \tilde{t}/\omega_{s}(\vec{q}))- \left\langle n_{s}(\vec{q}) \right\rangle}{\left\langle n_{s}(\vec{q})\right\rangle}\;,
\label{RenormN}
\end{equation}
which accounts for the fact that the equilibrium fluctuations of the occupation numbers are proportional
to their equilibrium value~$\left\langle n_{s}(\vec{q}) \right\rangle = {2\kb T}/\omega_s^2(\vec{q})$ and that
the natural time scale~$\tilde{t}$ of each mode is determined by its frequency~$\omega_{s}(\vec{q})$.
As a consequence, we inherently account for the typical $1/\omega^2$ dependence of the phonon lifetimes~\cite{Glassbrenner:1964fb,Garg:2011hi}
so that the respective ACFs~${G}[\Delta{n}_{s}(\vec{q},\tilde{t})]$ become
comparable and can be accurately interpolated in $\vec{q}$-space~(see Suppl. Mat.).

We validate this approach by applying it to FF simulation of Si based on
the Tersoff potential, which are known to be particularly challenging to convergence~\cite{He:2012hg,Howell:2012fh}. 
As shown in Fig.~\ref{ConvergenceSi} for 300 and 1000~K, the proposed interpolation
yields remarkable improvements with respect to size- and time-convergence:
Panel~(a) shows the dependence of~$\vec{\kappa}$ on the trajectory length, while panel~(b) shows the dependence 
of~$\vec{\kappa}$ on the supercell size. Compared to traditional brute force GK simulations, we achieve reliable
values for $\vec{\kappa}$ with sizes as small as 64 atoms and with trajectory lengths as short as 200~ps. This
translates 
into a computational speed-up of more 
than three orders of magnitude,
which in turn enables {\it ab initio} Green-Kubo calculations~(aiGK) with reasonable numerical effort.

\begin{figure}
  \centerline{\includegraphics[width=0.85\linewidth]{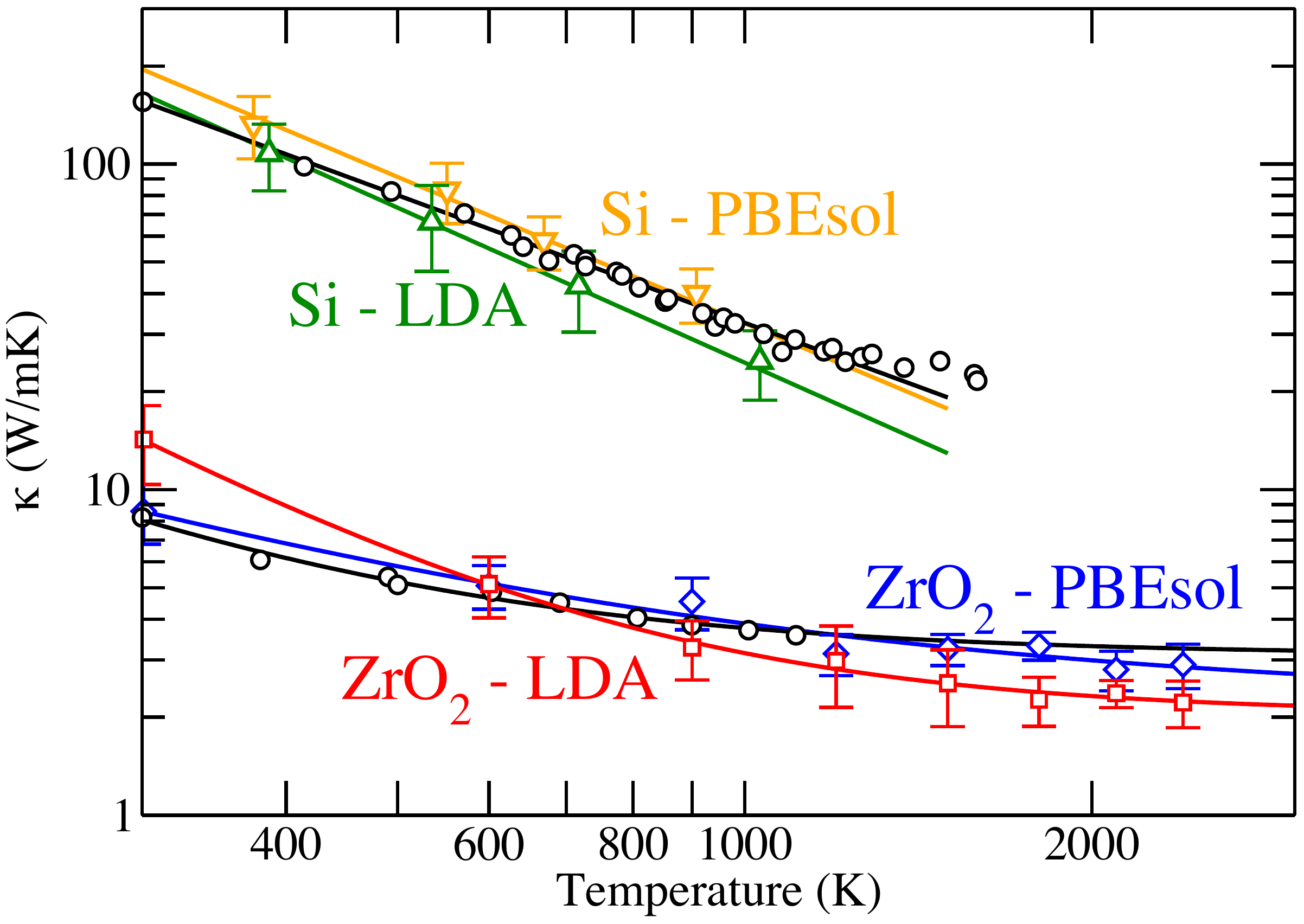}}
  \caption{Thermal conductivity ${\kappa}_{\alpha\alpha}$ computed using our aiGK method for Si~(LDA green and
   PBEsol orange triangles) and tetragonal ZrO$_2$~(LDA: red squares; PBEsol: blue diamonds). Lattice expansion 
   using the quasi-harmonic approximation and the tetragonal-cubic phase transition in ZrO$_2$~\cite{Carbogno:2014wa} 
   are taken into account. Tabulated values of~$\kappa$ are listed in the Suppl. Information. 
   Black circles denote experimental results for single crystals (or extrapolated
   to this limit) compiled from Ref.~\cite{Glassbrenner:1964fb,Bisson:2000tu,Youngblood:1988ws,Raghavan:1998va}. 
   The lines are generated by fitting the data with ${\kappa}(T)=a+b/T^c$. }
  \label{SiZrO2}
\end{figure}

\begin{figure}
  \centerline{\includegraphics[width=0.85\linewidth]{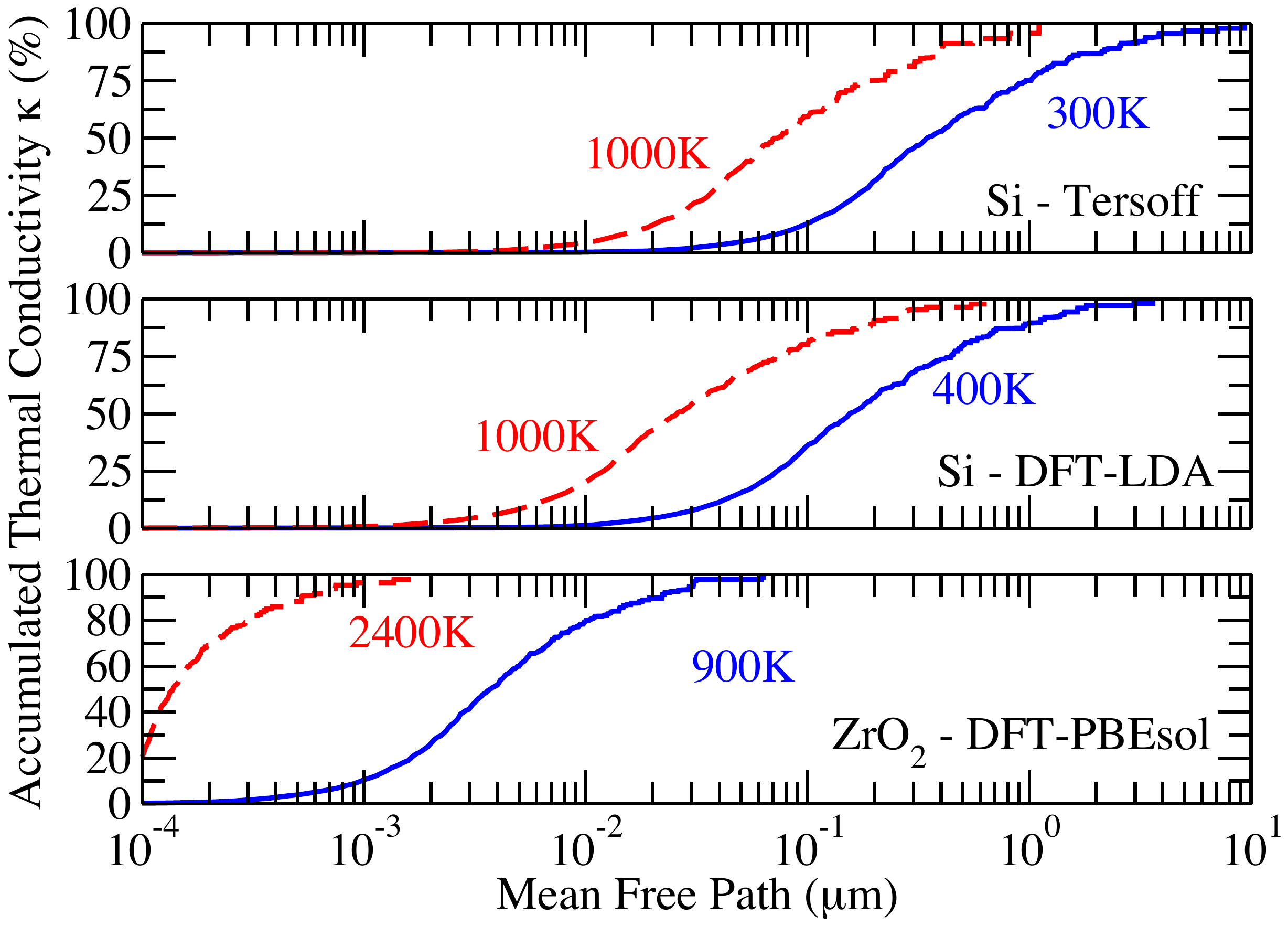}}
  \caption{Accumulated thermal conductivity versus mean free path for Si~(Tersoff/DFT-LDA) and ZrO$_2$~(DFT-PBEsol) at different temperatures.}
  \label{VACF}
\end{figure}

Next, we apply our aiGK technique to compute the temperature-dependent thermal conductivity of
Si~(64-atom cell) and tetragonal ZrO$_2$~(96-atom cell) both for the LDA and PBEsol functional. Six trajectories 
with a duration of 200~ps were used for each data point. In both cases~(cf.~Fig.~\ref{SiZrO2}),
we obtain good agreement with experimental data~\cite{Glassbrenner:1964fb,Bisson:2000tu,Youngblood:1988ws,Raghavan:1998va}.
For Si, we note that our aiGK extrapolation technique is responsible for up to 50\% of $\vec{\kappa}$, especially at 
low temperatures. Conversely, the thermal conductivities for ZrO$_2$ are mostly size- and time-converged within
the aiMD regime, so that corrections stemming from the extrapolation are always smaller than $10$\%. As the 
 accumulated thermal conductivities in Fig.~\ref{VACF} show, the reason for this behavior are the exceptionally
low phonon lifetimes or short mean free paths in ZrO$_2$. Interestingly, we can trace back 
this behavior to the peculiar, very anharmonic dynamics of zirconia at elevated temperatures.
Under such thermodynamic conditions, the oxygen atoms and the lattice tetragonality spontaneously re-orient
along a different cartesian direction in a switching mechanism~\cite{Carbogno:2014wa}.
These switches between different local minima of the potential-energy surface constitute
a severe violation of the harmonic approximation, given that the dynamics does no longer evolve around one minimum.
This is also reflected in the respective dependence on the functional: At the LDA level, the respective barriers are underestimated~\cite{Carbogno:2014wa},
so that $\vec{\kappa}$ decreases more drastically and at lower temperatures.

In summary, we presented an {\it ab initio} implementation of the Green-Kubo method that is applicable 
for the computation of thermal conductivities in solids, since the proposed unique definition of the heat
flux is derived from the virial theorem and based on the local stress tensor.
The developed extrapolation technique, which significantly lowers the required computational effort, enables 
to perform such computations within moderate time- and length scales for the aiMD trajectories. By this means,
we are able to accurately compute thermal conductivities for both extremely harmonic and anharmonic materials
on the same footing and at arbitrarily high temperatures. In particular, we are able to investigate materials
with very low thermal conductivities and high anharmonicities~(thermal barriers), for which perturbative treatments relying
on the approximate validity of the harmonic approximation would fail. Accordingly, the proposed technique enables
for the first time to perform accurate first-principles studies of such materials, which play a pivotal
role in a  multitude of scientific and technological applications,~e.g., as thermal barrier
coatings~\cite{Levi:2004vh} and thermoelectric elements~\cite{Snyder:2008dc}.

\begin{acknowledgments}
The authors thank J. Meyer, S. K. Estreicher, and C. G. Van de Walle for fruitful discussions. 
We are grateful to S. Lampenscherf and C. Levi for pointing out to us that a general, non-perturbative theory of heat transport was missing, so far. 
R. R. acknowledges the Alexander von Humboldt Foundation. 
The project received funding from the Einstein foundation~(project ETERNAL), the DFG cluster of excellence UNICAT, and 
the European Union's Horizon 2020 research and innovation program under grant agreement no. 676580 with The Novel Materials Discovery (NOMAD) Laboratory, a European Center of Excellence.
\end{acknowledgments}

\bibliography{bib2}

\begin{thebibliography}{34}
\expandafter\ifx\csname natexlab\endcsname\relax\def\natexlab#1{#1}\fi
\expandafter\ifx\csname bibnamefont\endcsname\relax
  \def\bibnamefont#1{#1}\fi
\expandafter\ifx\csname bibfnamefont\endcsname\relax
  \def\bibfnamefont#1{#1}\fi
\expandafter\ifx\csname citenamefont\endcsname\relax
  \def\citenamefont#1{#1}\fi
\expandafter\ifx\csname url\endcsname\relax
  \def\url#1{\texttt{#1}}\fi
\expandafter\ifx\csname urlprefix\endcsname\relax\def\urlprefix{URL }\fi
\providecommand{\bibinfo}[2]{#2}
\providecommand{\eprint}[2][]{\url{#2}}

\bibitem[{\citenamefont{Ashcroft and Mermin}(1976)}]{AshcroftMermin}
\bibinfo{author}{\bibfnamefont{N.~W.} \bibnamefont{Ashcroft}} \bibnamefont{and}
  \bibinfo{author}{\bibfnamefont{N.~D.} \bibnamefont{Mermin}},
  \emph{\bibinfo{title}{{Solid State Physics }}} (\bibinfo{publisher}{Saunders
  College Publishing}, \bibinfo{year}{1976}).

\bibitem[{\citenamefont{Donadio and Galli}(2009)}]{Donadio:2009fo}
\bibinfo{author}{\bibfnamefont{D.}~\bibnamefont{Donadio}} \bibnamefont{and}
  \bibinfo{author}{\bibfnamefont{G.}~\bibnamefont{Galli}},
  \bibinfo{journal}{Phys. Rev. Lett.} \textbf{\bibinfo{volume}{102}},
  \bibinfo{pages}{195901} (\bibinfo{year}{2009}).

\bibitem[{\citenamefont{Broido et~al.}(2005)\citenamefont{Broido, Ward, and
  Mingo}}]{Broido:2005kf}
\bibinfo{author}{\bibfnamefont{D.~A.} \bibnamefont{Broido}},
  \bibinfo{author}{\bibfnamefont{A.}~\bibnamefont{Ward}}, \bibnamefont{and}
  \bibinfo{author}{\bibfnamefont{N.}~\bibnamefont{Mingo}},
  \bibinfo{journal}{Phys. Rev. B} \textbf{\bibinfo{volume}{72}},
  \bibinfo{pages}{014308} (\bibinfo{year}{2005}).

\bibitem[{\citenamefont{Broido et~al.}(2007)\citenamefont{Broido, Malorny,
  Birner, Mingo, and Stewart}}]{Broido:2007ua}
\bibinfo{author}{\bibfnamefont{D.~A.} \bibnamefont{Broido}},
  \bibinfo{author}{\bibfnamefont{M.}~\bibnamefont{Malorny}},
  \bibinfo{author}{\bibfnamefont{G.}~\bibnamefont{Birner}},
  \bibinfo{author}{\bibfnamefont{N.}~\bibnamefont{Mingo}}, \bibnamefont{and}
  \bibinfo{author}{\bibfnamefont{D.~A.} \bibnamefont{Stewart}},
  \bibinfo{journal}{Appl. Phys. Lett.} \textbf{\bibinfo{volume}{91}},
  \bibinfo{pages}{231922} (\bibinfo{year}{2007}).

\bibitem[{\citenamefont{de~Koker}(2009)}]{deKoker:2009cr}
\bibinfo{author}{\bibfnamefont{N.}~\bibnamefont{de~Koker}},
  \bibinfo{journal}{Phys. Rev. Lett.} \textbf{\bibinfo{volume}{103}},
  \bibinfo{pages}{125902} (\bibinfo{year}{2009}).

\bibitem[{\citenamefont{Tang and Dong}(2010)}]{Tang:2010vu}
\bibinfo{author}{\bibfnamefont{X.}~\bibnamefont{Tang}} \bibnamefont{and}
  \bibinfo{author}{\bibfnamefont{J.}~\bibnamefont{Dong}},
  \bibinfo{journal}{Proc. Natl. Acad. Sci.}  (\bibinfo{year}{2010}).

\bibitem[{\citenamefont{Garg et~al.}(2011)\citenamefont{Garg, Bonini, Kozinsky,
  and Marzari}}]{Garg:2011hi}
\bibinfo{author}{\bibfnamefont{J.}~\bibnamefont{Garg}},
  \bibinfo{author}{\bibfnamefont{N.}~\bibnamefont{Bonini}},
  \bibinfo{author}{\bibfnamefont{B.}~\bibnamefont{Kozinsky}}, \bibnamefont{and}
  \bibinfo{author}{\bibfnamefont{N.}~\bibnamefont{Marzari}},
  \bibinfo{journal}{Phys. Rev. Lett.} \textbf{\bibinfo{volume}{106}},
  \bibinfo{pages}{045901} (\bibinfo{year}{2011}).

\bibitem[{\citenamefont{Ladd et~al.}(1986)\citenamefont{Ladd, Moran, and
  Hoover}}]{Ladd:1986tv}
\bibinfo{author}{\bibfnamefont{A.~J.~C.} \bibnamefont{Ladd}},
  \bibinfo{author}{\bibfnamefont{B.}~\bibnamefont{Moran}}, \bibnamefont{and}
  \bibinfo{author}{\bibfnamefont{W.~G.} \bibnamefont{Hoover}},
  \bibinfo{journal}{Physical Review B} \textbf{\bibinfo{volume}{34}},
  \bibinfo{pages}{5058} (\bibinfo{year}{1986}).

\bibitem[{\citenamefont{Turney et~al.}(2009)\citenamefont{Turney, Landry,
  McGaughey, and Amon}}]{Turney:2009ur}
\bibinfo{author}{\bibfnamefont{J.~E.}~\bibnamefont{Turney}},
  \bibinfo{author}{\bibfnamefont{E.~S.}~\bibnamefont{Landry}},
  \bibinfo{author}{\bibfnamefont{A.~J.~H.}~\bibnamefont{McGaughey}},
  \bibnamefont{and} \bibinfo{author}{\bibfnamefont{C.~H.}~\bibnamefont{Amon}},
  \bibinfo{journal}{Phys. Rev. B} \textbf{\bibinfo{volume}{79}},
  \bibinfo{pages}{064301} (\bibinfo{year}{2009}).

\bibitem[{\citenamefont{Gibbons and Estreicher}(2009)}]{Gibbons:2009hd}
\bibinfo{author}{\bibfnamefont{T.~M.} \bibnamefont{Gibbons}} \bibnamefont{and}
  \bibinfo{author}{\bibfnamefont{S.~K.} \bibnamefont{Estreicher}},
  \bibinfo{journal}{Phys. Rev. Lett.} \textbf{\bibinfo{volume}{102}},
  \bibinfo{pages}{255502} (\bibinfo{year}{2009}).

\bibitem[{\citenamefont{Gibbons et~al.}(2011)\citenamefont{Gibbons, Kang,
  Estreicher, and Carbogno}}]{Gibbons:2011je}
\bibinfo{author}{\bibfnamefont{T.~M.} \bibnamefont{Gibbons}},
  \bibinfo{author}{\bibfnamefont{B.}~\bibnamefont{Kang}},
  \bibinfo{author}{\bibfnamefont{S.~K.} \bibnamefont{Estreicher}},
  \bibnamefont{and} \bibinfo{author}{\bibfnamefont{C.}~\bibnamefont{Carbogno}},
  \bibinfo{journal}{Phys. Rev. B} \textbf{\bibinfo{volume}{84}},
  \bibinfo{pages}{035317} (\bibinfo{year}{2011}).

\bibitem[{\citenamefont{Stackhouse et~al.}(2010)\citenamefont{Stackhouse,
  Stixrude, and Karki}}]{Stackhouse:2010vg}
\bibinfo{author}{\bibfnamefont{S.}~\bibnamefont{Stackhouse}},
  \bibinfo{author}{\bibfnamefont{L.}~\bibnamefont{Stixrude}}, \bibnamefont{and}
  \bibinfo{author}{\bibfnamefont{B.~B.}~\bibnamefont{Karki}},
  \bibinfo{journal}{Phys. Rev. Lett.} \textbf{\bibinfo{volume}{104}},
  \bibinfo{pages}{208501} (\bibinfo{year}{2010}).

\bibitem[{\citenamefont{Tenenbaum et~al.}(1982)\citenamefont{Tenenbaum,
  Ciccotti, and Gallico}}]{Tenenbaum:1982wu}
\bibinfo{author}{\bibfnamefont{A.}~\bibnamefont{Tenenbaum}},
  \bibinfo{author}{\bibfnamefont{G.}~\bibnamefont{Ciccotti}}, \bibnamefont{and}
  \bibinfo{author}{\bibfnamefont{R.}~\bibnamefont{Gallico}},
  \bibinfo{journal}{Phys. Rev. A} \textbf{\bibinfo{volume}{25}},
  \bibinfo{pages}{2778} (\bibinfo{year}{1982}).

\bibitem[{\citenamefont{Schelling et~al.}(2002)\citenamefont{Schelling,
  Phillpot, and Keblinski}}]{Schelling:2002jl}
\bibinfo{author}{\bibfnamefont{P.~K.}~\bibnamefont{Schelling}},
  \bibinfo{author}{\bibfnamefont{S.~R.}~\bibnamefont{Phillpot}}, \bibnamefont{and}
  \bibinfo{author}{\bibfnamefont{P.}~\bibnamefont{Keblinski}},
  \bibinfo{journal}{Phys. Rev. B} \textbf{\bibinfo{volume}{65}},
  \bibinfo{pages}{144306} (\bibinfo{year}{2002}).

\bibitem[{\citenamefont{He et~al.}(2012)\citenamefont{He, Savi{\'c}, Donadio,
  and Galli}}]{He:2012hg}
\bibinfo{author}{\bibfnamefont{Y.}~\bibnamefont{He}},
  \bibinfo{author}{\bibfnamefont{I.}~\bibnamefont{Savi{\'c}}},
  \bibinfo{author}{\bibfnamefont{D.}~\bibnamefont{Donadio}}, \bibnamefont{and}
  \bibinfo{author}{\bibfnamefont{G.}~\bibnamefont{Galli}},
  \bibinfo{journal}{Phys. Chem. Chem. Phys.} \textbf{\bibinfo{volume}{14}},
  \bibinfo{pages}{16209} (\bibinfo{year}{2012}).

\bibitem[{\citenamefont{Kubo et~al.}(1957)\citenamefont{Kubo, Yokota, and
  Nakajima}}]{Kubo:1957tf}
\bibinfo{author}{\bibfnamefont{R.}~\bibnamefont{Kubo}},
  \bibinfo{author}{\bibfnamefont{M.}~\bibnamefont{Yokota}}, \bibnamefont{and}
  \bibinfo{author}{\bibfnamefont{S.}~\bibnamefont{Nakajima}},
  \bibinfo{journal}{J. Phys. Soc. Japan} \textbf{\bibinfo{volume}{12}},
  \bibinfo{pages}{1203} (\bibinfo{year}{1957}).

\bibitem[{\citenamefont{MacDowell}(1999)}]{MACDOWELL:1999fe}
\bibinfo{author}{\bibfnamefont{L.~G.} \bibnamefont{MacDowell}},
  \bibinfo{journal}{Mol. Phys.} \textbf{\bibinfo{volume}{96}},
  \bibinfo{pages}{881} (\bibinfo{year}{1999}).

\bibitem[{\citenamefont{Hardy}(1963)}]{Hardy:1963gh}
\bibinfo{author}{\bibfnamefont{R.~J.} \bibnamefont{Hardy}},
  \bibinfo{journal}{Phys Rev} \textbf{\bibinfo{volume}{132}},
  \bibinfo{pages}{168} (\bibinfo{year}{1963}).

\bibitem[{\citenamefont{Helfand}(1960)}]{Helfand:1960uy}
\bibinfo{author}{\bibfnamefont{E.}~\bibnamefont{Helfand}},
  \bibinfo{journal}{Phys Rev} \textbf{\bibinfo{volume}{119}},
  \bibinfo{pages}{1} (\bibinfo{year}{1960}).

\bibitem[{\citenamefont{Yu et~al.}(2011)\citenamefont{Yu, Trinkle, and
  Martin}}]{Yu:2011kv}
\bibinfo{author}{\bibfnamefont{M.}~\bibnamefont{Yu}},
  \bibinfo{author}{\bibfnamefont{D.~R.}~\bibnamefont{Trinkle}}, \bibnamefont{and}
  \bibinfo{author}{\bibfnamefont{R.~M.}~\bibnamefont{Martin}},
  \bibinfo{journal}{Phys. Rev. B} \textbf{\bibinfo{volume}{83}},
  \bibinfo{pages}{115113} (\bibinfo{year}{2011}).

\bibitem[{\citenamefont{Marcolongo et~al.}(2016)\citenamefont{Marcolongo,
  Umari, and Baroni}}]{Marcolongo:2016dn}
\bibinfo{author}{\bibfnamefont{A.}~\bibnamefont{Marcolongo}},
  \bibinfo{author}{\bibfnamefont{P.}~\bibnamefont{Umari}}, \bibnamefont{and}
  \bibinfo{author}{\bibfnamefont{S.}~\bibnamefont{Baroni}},
  \bibinfo{journal}{Nat. Phys.} \textbf{\bibinfo{volume}{12}},
  \bibinfo{pages}{80} (\bibinfo{year}{2016}).

\bibitem[{\citenamefont{Howell}(2012)}]{Howell:2012fh}
\bibinfo{author}{\bibfnamefont{P.~C.} \bibnamefont{Howell}},
  \bibinfo{journal}{J. Chem. Phys.} \textbf{\bibinfo{volume}{137}},
  \bibinfo{pages}{224111} (\bibinfo{year}{2012}).

\bibitem[{\citenamefont{Knuth et~al.}(2015)\citenamefont{Knuth, Carbogno,
  Atalla, Blum, and Scheffler}}]{Knuth:2015kc}
\bibinfo{author}{\bibfnamefont{F.}~\bibnamefont{Knuth}},
  \bibinfo{author}{\bibfnamefont{C.}~\bibnamefont{Carbogno}},
  \bibinfo{author}{\bibfnamefont{V.}~\bibnamefont{Atalla}},
  \bibinfo{author}{\bibfnamefont{V.}~\bibnamefont{Blum}}, \bibnamefont{and}
  \bibinfo{author}{\bibfnamefont{M.}~\bibnamefont{Scheffler}},
  \bibinfo{journal}{Comp. Phys. Comm.} \textbf{\bibinfo{volume}{190}},
  \bibinfo{pages}{33} (\bibinfo{year}{2015}).

\bibitem[{\citenamefont{Blum et~al.}(2009)\citenamefont{Blum, Gehrke, Hanke,
  Havu, Havu, Ren, Reuter, and Scheffler}}]{Blum:2009fe}
\bibinfo{author}{\bibfnamefont{V.}~\bibnamefont{Blum}},
  \bibinfo{author}{\bibfnamefont{R.}~\bibnamefont{Gehrke}},
  \bibinfo{author}{\bibfnamefont{F.}~\bibnamefont{Hanke}},
  \bibinfo{author}{\bibfnamefont{P.}~\bibnamefont{Havu}},
  \bibinfo{author}{\bibfnamefont{V.}~\bibnamefont{Havu}},
  \bibinfo{author}{\bibfnamefont{X.}~\bibnamefont{Ren}},
  \bibinfo{author}{\bibfnamefont{K.}~\bibnamefont{Reuter}}, \bibnamefont{and}
  \bibinfo{author}{\bibfnamefont{M.}~\bibnamefont{Scheffler}},
  \bibinfo{journal}{Comp. Phys. Comm.} \textbf{\bibinfo{volume}{180}},
  \bibinfo{pages}{2175} (\bibinfo{year}{2009}).

\bibitem[{\citenamefont{Peierls}(1929)}]{Peierls:1929jv}
\bibinfo{author}{\bibfnamefont{R.}~\bibnamefont{Peierls}},
  \bibinfo{journal}{Ann. Phys.} \textbf{\bibinfo{volume}{395}},
  \bibinfo{pages}{1055} (\bibinfo{year}{1929}).

\bibitem[{\citenamefont{McGaughey and Kaviany}(2004)}]{McGaughey:2004iz}
\bibinfo{author}{\bibfnamefont{A.~J.~H.}~\bibnamefont{McGaughey}} \bibnamefont{and}
  \bibinfo{author}{\bibfnamefont{M.}~\bibnamefont{Kaviany}},
  \bibinfo{journal}{Phys. Rev. B} \textbf{\bibinfo{volume}{69}},
  \bibinfo{pages}{094303} (\bibinfo{year}{2004}).

\bibitem[{\citenamefont{Parlinski et~al.}(1997)\citenamefont{Parlinski, Li, and
  Kawazoe}}]{Parlinski:1997kr}
\bibinfo{author}{\bibfnamefont{K.}~\bibnamefont{Parlinski}},
  \bibinfo{author}{\bibfnamefont{Z.-Q.} \bibnamefont{Li}}, \bibnamefont{and}
  \bibinfo{author}{\bibfnamefont{Y.}~\bibnamefont{Kawazoe}},
  \bibinfo{journal}{Phys. Rev. Lett.} \textbf{\bibinfo{volume}{78}},
  \bibinfo{pages}{4063} (\bibinfo{year}{1997}).

\bibitem[{\citenamefont{Glassbrenner and Slack}(1964)}]{Glassbrenner:1964fb}
\bibinfo{author}{\bibfnamefont{C.~J.} \bibnamefont{Glassbrenner}}
  \bibnamefont{and} \bibinfo{author}{\bibfnamefont{G.~A.} \bibnamefont{Slack}},
  \bibinfo{journal}{Phys Rev} \textbf{\bibinfo{volume}{134}},
  \bibinfo{pages}{A1058} (\bibinfo{year}{1964}).

\bibitem[{\citenamefont{Carbogno et~al.}(2014)\citenamefont{Carbogno, Levi,
  Van~de Walle, and Scheffler}}]{Carbogno:2014wa}
\bibinfo{author}{\bibfnamefont{C.}~\bibnamefont{Carbogno}},
  \bibinfo{author}{\bibfnamefont{C.~G.} \bibnamefont{Levi}},
  \bibinfo{author}{\bibfnamefont{C.~G.} \bibnamefont{Van~de Walle}},
  \bibnamefont{and}
  \bibinfo{author}{\bibfnamefont{M.}~\bibnamefont{Scheffler}},
  \bibinfo{journal}{Physical Review B} \textbf{\bibinfo{volume}{90}},
  \bibinfo{pages}{144109} (\bibinfo{year}{2014}).

\bibitem[{\citenamefont{Bisson et~al.}(2000)\citenamefont{Bisson, Fournier,
  Poulain, Lavigne, and M{\'e}vrel}}]{Bisson:2000tu}
\bibinfo{author}{\bibfnamefont{J.-F.} \bibnamefont{Bisson}},
  \bibinfo{author}{\bibfnamefont{D.}~\bibnamefont{Fournier}},
  \bibinfo{author}{\bibfnamefont{M.}~\bibnamefont{Poulain}},
  \bibinfo{author}{\bibfnamefont{O.}~\bibnamefont{Lavigne}}, \bibnamefont{and}
  \bibinfo{author}{\bibfnamefont{R.}~\bibnamefont{M{\'e}vrel}},
  \bibinfo{journal}{J. Am. Cer. Soc.} \textbf{\bibinfo{volume}{83}},
  \bibinfo{pages}{1993} (\bibinfo{year}{2000}).

\bibitem[{\citenamefont{Youngblood et~al.}(1988)\citenamefont{Youngblood, Rice,
  and Ingel}}]{Youngblood:1988ws}
\bibinfo{author}{\bibfnamefont{G.~E.} \bibnamefont{Youngblood}},
  \bibinfo{author}{\bibfnamefont{R.~W.} \bibnamefont{Rice}}, \bibnamefont{and}
  \bibinfo{author}{\bibfnamefont{R.~P.} \bibnamefont{Ingel}},
  \bibinfo{journal}{J. Am. Cer. Soc.} \textbf{\bibinfo{volume}{71}},
  \bibinfo{pages}{255} (\bibinfo{year}{1988}).

\bibitem[{\citenamefont{Raghavan et~al.}(1998)\citenamefont{Raghavan, Wang,
  Dinwiddie, Porter, and Mayo}}]{Raghavan:1998va}
\bibinfo{author}{\bibfnamefont{S.}~\bibnamefont{Raghavan}},
  \bibinfo{author}{\bibfnamefont{H.}~\bibnamefont{Wang}},
  \bibinfo{author}{\bibfnamefont{R.}~\bibnamefont{Dinwiddie}},
  \bibinfo{author}{\bibfnamefont{W.}~\bibnamefont{Porter}}, \bibnamefont{and}
  \bibinfo{author}{\bibfnamefont{M.}~\bibnamefont{Mayo}},
  \bibinfo{journal}{Scripta Materialia} \textbf{\bibinfo{volume}{39}},
  \bibinfo{pages}{1119} (\bibinfo{year}{1998}).

\bibitem[{\citenamefont{Levi}(2004)}]{Levi:2004vh}
\bibinfo{author}{\bibfnamefont{C.~G.} \bibnamefont{Levi}},
  \bibinfo{journal}{Curr. Opin. Solid State Mater. Sci.}
  \textbf{\bibinfo{volume}{8}}, \bibinfo{pages}{77} (\bibinfo{year}{2004}).

\bibitem[{\citenamefont{Snyder and Toberer}(2008)}]{Snyder:2008dc}
\bibinfo{author}{\bibfnamefont{G.~J.} \bibnamefont{Snyder}} \bibnamefont{and}
  \bibinfo{author}{\bibfnamefont{E.~S.} \bibnamefont{Toberer}},
  \bibinfo{journal}{Nature Materials} \textbf{\bibinfo{volume}{7}},
  \bibinfo{pages}{105} (\bibinfo{year}{2008}).

\end{thebibliography}

\end{document}